\documentclass[10pt,journal]{IEEEtran}

\ifx\pdfoutput\undefined
\usepackage{graphicx}
\graphicspath{{eps_figs/}}
\else
\usepackage[pdftex]{graphicx}
\graphicspath{{pdf_figs/}}
\fi

\usepackage{latexsym,amssymb,amsmath,cite,ifthen,afterpage}

\newcommand{\iftwocolumn}[2]{\ifthenelse{\boolean{@twocolumn}}{#1}{#2}}

\newcommand{\Fig}[1]{Fig.~\ref{#1}}

\newcommand{\andor}{\,/\,}
\newcommand{\eqdef}{\stackrel{\scriptscriptstyle\bigtriangleup}{=} }

\newcommand{\smpl}[2]{#1^{(#2)}}
\newcommand{\cond}{\hspace{0.02em}|\hspace{0.08em}}

\newcommand{\R}{\mathbb{R}}

\newcommand{\Y}{{\bf Y}} 
\newcommand{\X}{{\bf X}}

\newcommand{\y}{{\bf y}}
\newcommand{\x}{{\bf x}}

\newcommand{\vct}[1]{\mathbf{#1}}    

\newcommand{\calX}{\mathcal{X}}

\newcommand{\calS}{\mathcal{S}}
\newcommand{\0}{{\bf 0}}

\newcommand{\E}{\operatorname{E}}

\newcommand{\msgf}[2]{\protect\overrightarrow{#1}_{\!#2}}
\newcommand{\msgb}[2]{\protect\overleftarrow{#1}_{\!#2}}

\newcounter{examplecntr}
{\begin{trivlist}\small\item[]\refstepcounter{examplecntr}%
 {\bfseries Example~\theexamplecntr%
  \ifthenelse{\equal{#1}{}}{}{ (#1)}.
}}%
{\end{trivlist}}

\newcounter{theoremcntr}
{\begin{trivlist}\item[]\refstepcounter{theoremcntr}%
{\bfseries Theorem~\thetheoremcntr%
  \ifthenelse{\equal{#1}{}}{}{ (#1)}.
}}%
{\hfill$\Box$\end{trivlist}}

\newcommand{\cent}[1]{\makebox(0,0){#1}}
\newcommand{\pos}[2]{\makebox(0,0)[#1]{#2}}

\begin{document}
\DeclareGraphicsExtensions{.pdf}


\title{Monte Carlo Algorithms\\ for the Partition Function and Information Rates\\ 
of Two-Dimensional Channels}


\author{%
Mehdi Molkaraie,~\IEEEmembership{Member,~IEEE} and 
Hans-Andrea Loeliger,~\IEEEmembership{Fellow,~IEEE}
\thanks{%
Mehdi Molkaraie was with 
the Dept.\ of Information Technology and Electrical Engineering, 
ETH Z\"urich, CH-8092 Z\"urich, Switzerland. He is now with the
Dept.\ of Statistics and Actuarial Science, University of Waterloo,
Waterloo N2L 3G1, Canada. Email: mmolkaraie@uwaterloo.ca.
Hans-Andrea Loeliger is with 
the Dept.\ of Information Technology and Electrical Engineering, 
ETH Z\"urich, CH-8092 Z\"urich, Switzerland.
Email: loeliger@isi.ee.ethz.ch.

Preliminary versions of the material of this paper were presented
in \cite{LoMo:ISIT2008c,LoMo:ITW2009,MoLo:ISIT2010}.
}
}

\maketitle

\begin{abstract}
The paper proposes Monte Carlo algorithms for 
the computation of the information rate 
of two-dimensional source\andor{}channel models. 
The focus of the paper is on binary-input channels with 
constraints on the allowed input configurations. 
The problem of numerically computing the information rate, 
and even the noiseless capacity, of such channels
has so far remained largely unsolved.
Both problems can be reduced to computing 
a Monte Carlo estimate of a partition function.
The proposed algorithms use tree-based Gibbs sampling 
and multilayer (multitemperature) importance sampling.
The viability of the proposed algorithms 
is demonstrated by simulation results.
\end{abstract}


\section{Introduction}
\label{sec:Introduction}

Numerically computing the Shannon information rate 
for source\andor{}channel models with memory can be difficult. 
In many cases of practical interest, 
analytical results are not available or hard to evaluate numerically.
For a large class of channels, however, Monte Carlo methods 
as proposed in 
\cite{ArLg:ICC2001c,ALVKZ:sbcir2006,PSS:2001}
have been shown to yield reliable numerical results.

In this paper, we consider the extension of such Monte Carlo methods 
to source\andor{}channel models with a two-dimensional (\mbox{2-D}) structure. 
The focus of the paper is on 2-D binary-input channels with constraints 
on the allowed input configurations;
for example, we consider the channel where adjacent 
channel input symbols must not both have the value~1. 
Variations of such channels are of interest 
in magnetic and optical storage, where the constraints are imposed,
e.g., in order to reduce the intersymbol interference or to help in
timing control \cite{Sch:04, Roth:97, BVEK:05, ISW:98, WKKM:09}. 
We will consider both noiseless and noisy versions of such channels.
With suitable modifications
(simplifications),
the methods of this paper can also be applied 
to other \mbox{2-D} source\andor{}channel models 
such as channels with intersymbol interference.

In the one-dimensional (\mbox{1-D}) case, computing the capacity of noiseless constrained channels 
was addressed and solved by Shannon \cite{Sh:48}, see also \cite{Sch:04}. 
For the noisy case, 
the Monte Carlo methods of \cite{ArLg:ICC2001c,ALVKZ:sbcir2006,PSS:2001} can be used to compute the information rate. 
The 2-D case is harder. 
Even the noiseless capacity is hard to compute numerically:
while very tight analytical results are available 
for a number of special cases (e.g., \cite{CW:98,WB:98,KZ:99,IKNZ:00,CE:06,TR:11}),
other cases have remained open problems.
The noisy case has remained largely unsolved. 

The capacity of a noiseless constrained channel 
is essentially 
the logarithm of the partition function of the corresponding indicator function 
(see Section~\ref{sec:PartitionFunction}). 
Moreover,
computing the information rate of noisy source\andor{}channel models
can also be reduced to the computation of a partition function 
(see Section~\ref{sec:EstInfo}). 
The heart of the paper, therefore, are
Monte Carlo algorithms for the computation of partition functions. 
Several such algorithms are well known 
\cite{OgTa:eip1981,JS:93,PoGo:sapf1997}, see also~\cite{MK:mct1998,Bre:99},
but some modifications will be necessary for the problems of interest in this paper.
In particular, we will find tree-based Gibbs sampling 
(due to Hamze and de Freitas \cite{FF:f2t2004c}) extremely useful.
We will observe that Monte Carlo estimates of a partition function 
may actually be obtained as a by-product of tree-based Gibbs sampling, 
which does not seem to have been noticed before.

In related prior work, Monte Carlo algorithms 
have been used to compute bounds on, or approximations of, 
the information rate of 2-D source\andor{}channels with memory 
\cite{SSSKWW:2DChannels2008,CS:06}.
Some of this work uses generalized belief propagation \cite{YFW:05},
which appears to yield very good approximations to the information rate 
\cite{SSSKWW:2DChannels2008},
\cite{SaMo:ISIT2010,MoLo:ISIT2010,SaMo:2011}.

%
%

In contrast to all this prior work, 
the Monte Carlo methods of this paper are asymptotically unbiased, 
i.e., 
in the limit of infinitely many samples, 
the estimates are guaranteed to converge to the desired quantity 
(the information rate). 
Moreover, the focus of this paper is on constrained channels, 
for which these computational problems are harder 
than for intersymbol interference channels 
(cf.\ Section~\ref{sec:EstInfoMethod}). 

The empirical success of the proposed algorithms is epitomized 
by \Fig{fig:SNR2}, which shows the uniform-input information rate 
of a binary-input channel with input constraints 
and additive white Gaussian noise (AWGN). 
As far as known to the authors, 
no such figure (for such a channel) has been presented before.


If the reader is not familiar with Gibbs sampling, the following comments on the 
general nature of this work may be in order.
First, Gibbs sampling is easily proved (under very mild conditions) to yield samples that 
are \emph{eventually} distributed 
according to the desired distribution and \emph{asymptotically} independent~\cite{Bre:99} (i.e., deleting 
the first $t$ samples and decimating the remaining sample sequence by a factor $m$ results in an i.i.d.\ sequence 
in the limit $t, m\rightarrow\infty$). However, the dependencies among the samples may decay extremely slowly, 
which is the pivotal issue with Gibbs sampling and makes naive Gibbs sampling perfectly useless for the 
problems of this paper (and for many other problems). The challenge, therefore, is to speed up Gibbs 
sampling (i.e., to decrease the dependencies of the samples) by various additional tricks and insights so 
that it becomes useful.


Second, the class of problems 
for which the methods proposed in this paper will work is not easily expressed in exact 
terms. Again, the issue is not formal applicability (which is quite sweeping), but the 
speed of convergence, which strongly depends on the particulars of the case
and is not easily predicted.

The paper is organized as follows. 
In Section~\ref{sec:PartitionFunction}, we review partition functions 
and noiseless 2-D constrained channels, 
and we introduce the corresponding notation. 
In Section~\ref{sec:BasicMonteCarlo}, we review several Monte Carlo 
algorithms that will be used in this paper.
However, additional ideas are necessary to make these algorithms work 
for our applications. 
In particular, we will need tree-based Gibbs sampling as described 
in Section~\ref{sec:TBGZ}. 
The application to noiseless constrained channels is 
demonstrated in Section~\ref{sec:NoiselessCapNumEx}. 
The application to noisy channels is described and demonstrated in Section~\ref{sec:EstInfo}. 
The appendix reviews sampling from Markov chains, which is needed in 
Section~\ref{sec:TBGZ}.

\section{Partition Function of 2-D Graphical Models}
\label{sec:PartitionFunction}


Let $\calX_1, \calX_2, \ldots, \calX_N$ be finite sets,
let $\calX$ be the Cartesian product 
$\calX \eqdef \calX_1 \times \calX_2 \times \ldots \times \calX_N$,
and let $f$ be a nonnegative function
$f: \calX \rightarrow \R$.
We are interested in computing (exactly or approximately) the 
\emph{partition function} 

\begin{equation} \label{eqn:PartitionFunction}
Z_f \eqdef \sum_{\x\in \calX} f(\x)
\end{equation}
for cases where 
\begin{itemize}
\item
$N$ is large and
\item
$f$ has a suitable factorization (as will be detailed below).
\end{itemize}
We will usually assume $Z_f \neq 0$. Then
\begin{equation} \label{eqn:Pdef}
p_f(\x) \eqdef \frac{f(\x)}{Z_f}
\end{equation}
is a probability mass function on $\calX$.
We also define the support of $f$ (and of $p_f$) as 
\begin{equation} \label{eqn:Admiss}
\calS_f \eqdef \{ \x\in\calX: f(\x) > 0 \}.
\end{equation}

If $f(\x)$ has a cycle-free factor graph representation 
(and if $|\calX_1|, |\calX_2|, \ldots, |\calX_N|$ are not too large), 
then $Z_f$ can be computed 
efficiently by sum-product message passing \cite{Lg:ifg2004,KFL:fg2000}.
In this paper, however, we consider 
factor graphs with cycles. 
In particular, we are interested in examples of the following type. 

\begin{figure}[t]
\setlength{\unitlength}{0.94mm}
\centering
\begin{picture}(76,64)(0,0)
\small
%
\put(0,60){\framebox(4,4){$=$}}
\put(4,62){\line(1,0){8}} \put(8,63){\pos{bc}{$X_1$}}
\put(12,60){\framebox(4,4){}}
\put(16,62){\line(1,0){8}}
\put(24,60){\framebox(4,4){$=$}}
\put(28,62){\line(1,0){8}}  \put(32,63){\pos{bc}{$X_2$}}
\put(36,60){\framebox(4,4){}}
\put(40,62){\line(1,0){8}}
\put(48,60){\framebox(4,4){$=$}}
\put(52,62){\line(1,0){8}} \put(56,63){\pos{bc}{$X_3$}}
\put(60,60){\framebox(4,4){}}
\put(64,62){\line(1,0){8}}
\put(72,60){\framebox(4,4){$=$}}
\put(2,54){\line(0,1){6}}
\put(0,50){\framebox(4,4){}}
\put(2,50){\line(0,-1){6}}
\put(26,54){\line(0,1){6}}
\put(24,50){\framebox(4,4){}}
\put(26,50){\line(0,-1){6}}
\put(50,54){\line(0,1){6}}
\put(48,50){\framebox(4,4){}}
\put(50,50){\line(0,-1){6}}
\put(74,54){\line(0,1){6}}
\put(72,50){\framebox(4,4){}}
\put(74,50){\line(0,-1){6}}
\put(0,40){\framebox(4,4){$=$}}
\put(4,42){\line(1,0){8}}
\put(12,40){\framebox(4,4){}}
\put(16,42){\line(1,0){8}}
\put(24,40){\framebox(4,4){$=$}}
\put(28,42){\line(1,0){8}}
\put(36,40){\framebox(4,4){}}
\put(40,42){\line(1,0){8}}
\put(48,40){\framebox(4,4){$=$}}
\put(52,42){\line(1,0){8}}
\put(60,40){\framebox(4,4){}}
\put(64,42){\line(1,0){8}}
\put(72,40){\framebox(4,4){$=$}}
\put(2,34){\line(0,1){6}}
\put(0,30){\framebox(4,4){}}
\put(2,30){\line(0,-1){6}}
\put(26,34){\line(0,1){6}}
\put(24,30){\framebox(4,4){}}
\put(26,30){\line(0,-1){6}}
\put(50,34){\line(0,1){6}}
\put(48,30){\framebox(4,4){}}
\put(50,30){\line(0,-1){6}}
\put(74,34){\line(0,1){6}}
\put(72,30){\framebox(4,4){}}
\put(74,30){\line(0,-1){6}}
\put(0,20){\framebox(4,4){$=$}}
\put(4,22){\line(1,0){8}}
\put(12,20){\framebox(4,4){}}
\put(16,22){\line(1,0){8}}
\put(24,20){\framebox(4,4){$=$}}
\put(28,22){\line(1,0){8}}
\put(36,20){\framebox(4,4){}}
\put(40,22){\line(1,0){8}}
\put(48,20){\framebox(4,4){$=$}}
\put(52,22){\line(1,0){8}}
\put(60,20){\framebox(4,4){}}
\put(64,22){\line(1,0){8}}
\put(72,20){\framebox(4,4){$=$}}
\put(2,14){\line(0,1){6}}
\put(0,10){\framebox(4,4){}}
\put(2,10){\line(0,-1){6}}
\put(26,14){\line(0,1){6}}
\put(24,10){\framebox(4,4){}}
\put(26,10){\line(0,-1){6}}
\put(50,14){\line(0,1){6}}
\put(48,10){\framebox(4,4){}}
\put(50,10){\line(0,-1){6}}
\put(74,14){\line(0,1){6}}
\put(72,10){\framebox(4,4){}}
\put(74,10){\line(0,-1){6}}
\put(0,0){\framebox(4,4){$=$}}
\put(4,2){\line(1,0){8}}
\put(12,0){\framebox(4,4){}}
\put(16,2){\line(1,0){8}}
\put(24,0){\framebox(4,4){$=$}}
\put(28,2){\line(1,0){8}}
\put(36,0){\framebox(4,4){}}
\put(40,2){\line(1,0){8}}
\put(48,0){\framebox(4,4){$=$}}
\put(52,2){\line(1,0){8}}
\put(60,0){\framebox(4,4){}}
\put(64,2){\line(1,0){8}}
\put(72,0){\framebox(4,4){$=$}}
\end{picture}
\caption{\label{fig:2DGrid}%
Forney factor graph 
of the indicator function (\ref{eqn:NoAdjacentOnesFunction}). 
The unlabeled boxes represent factors as in~(\ref{eqn:NoAdjacentOnesFactor}).
}
\end{figure}
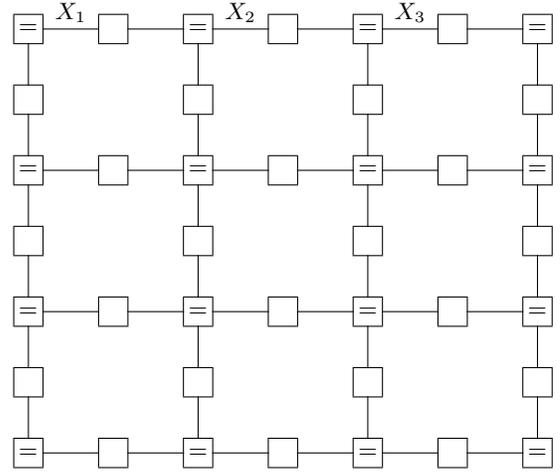

\begin{trivlist}
\item\textbf{Example: Simple 2-D Constrained Channel}\\
Consider a grid of $N=M\times M$ binary 
(i.e., $\{0,1\}$-valued) 
variables 
$x_1,\ldots,x_N$
with the constraint that no two (horizontally or vertically) adjacent 
variables have both the value~$1$. 
Let $f: \{0,1\}^N \rightarrow \{0,1\}$ 
be the indicator function 
of this constraint, which can be factored into 
\begin{equation} \label{eqn:NoAdjacentOnesFunction}
f(x_1,\ldots,x_N) = \prod_{\text{$k,\ell$ adjacent}} \kappa(x_k,x_\ell),
\end{equation}
where the product runs over all adjacent pairs $(k,\ell)$ 
and with factors
\begin{equation} \label{eqn:NoAdjacentOnesFactor}
\kappa(x_k,x_\ell) = \left\{ \begin{array}{ll}
     0, & \text{if $x_k = x_\ell = 1$} \\
     1, & \text{otherwise.}
  \end{array} \right.
\end{equation}
The corresponding Forney factor graph of $f$ is shown in \Fig{fig:2DGrid},
where the boxes labeled ``$=$'' are equality constraints \cite{Lg:ifg2004}. 
(Note that, in Forney factor graphs, variables are represented by edges. \Fig{fig:2DGrid} 
may also be viewed as a factor graph as in \cite{KFL:fg2000} 
where the boxes labeled ``$=$'' are the variable nodes.)

Note that, in this example, $Z_f=|\calS_f|$.

This example is known as the 2-D $(1,\infty)$ run-length limited constrained channel~\cite{Sch:04}.
%
%
The quantity 
\begin{equation} \label{eqn:CM}
C_M \eqdef \frac{1}{N}\log_2{Z_f}
\end{equation}
is known as the (finite-size) noiseless capacity of the channel.

For this particular example, upper and lower bounds on 
the infinite-size noiseless capacity
\begin{equation}
C_\infty \eqdef \lim_{M\rightarrow\infty} C_M
\end{equation}
were first proposed in~\cite{CW:98} and improved in~\cite{WB:98} and~\cite{NZ:2000}. 
The bounds in~\cite{NZ:2000} agree on nine decimal digits,
which far exceeds the accuracy that can be achieved with the Monte Carlo methods 
of the present paper. 
However, the methods proposed in this paper work also for various generalizations 
of this example for which 
no tight bounds are known.
\hfill$\Box$
\end{trivlist}


\newcommand{\drawgrid}{%
\begin{picture}(76,64)(0,0)
\small
%
\put(0,60){\framebox(4,4){$=$}}
\put(4,62){\line(1,0){8}} 
\put(12,60){\framebox(4,4){}}
\put(16,62){\line(1,0){8}}
\put(24,60){\framebox(4,4){$=$}}
\put(28,62){\line(1,0){8}}
\put(36,60){\framebox(4,4){}}
\put(40,62){\line(1,0){8}}
\put(48,60){\framebox(4,4){$=$}}
\put(52,62){\line(1,0){8}}
\put(60,60){\framebox(4,4){}}
\put(64,62){\line(1,0){8}}
\put(72,60){\framebox(4,4){$=$}}
\put(2,54){\line(0,1){6}}
\put(0,50){\framebox(4,4){}}
\put(2,50){\line(0,-1){6}}
\put(26,54){\line(0,1){6}}
\put(24,50){\framebox(4,4){}}
\put(26,50){\line(0,-1){6}}
\put(50,54){\line(0,1){6}}
\put(48,50){\framebox(4,4){}}
\put(50,50){\line(0,-1){6}}
\put(74,54){\line(0,1){6}}
\put(72,50){\framebox(4,4){}}
\put(74,50){\line(0,-1){6}}
\put(0,40){\framebox(4,4){$=$}}
\put(4,42){\line(1,0){8}}
\put(12,40){\framebox(4,4){}}
\put(16,42){\line(1,0){8}}
\put(24,40){\framebox(4,4){$=$}}
\put(28,42){\line(1,0){8}}
\put(36,40){\framebox(4,4){}}
\put(40,42){\line(1,0){8}}
\put(48,40){\framebox(4,4){$=$}}
\put(52,42){\line(1,0){8}}
\put(60,40){\framebox(4,4){}}
\put(64,42){\line(1,0){8}}
\put(72,40){\framebox(4,4){$=$}}
\put(2,34){\line(0,1){6}}
\put(0,30){\framebox(4,4){}}
\put(2,30){\line(0,-1){6}}
\put(26,34){\line(0,1){6}}
\put(24,30){\framebox(4,4){}}
\put(26,30){\line(0,-1){6}}
\put(50,34){\line(0,1){6}}
\put(48,30){\framebox(4,4){}}
\put(50,30){\line(0,-1){6}}
\put(74,34){\line(0,1){6}}
\put(72,30){\framebox(4,4){}}
\put(74,30){\line(0,-1){6}}
\put(0,20){\framebox(4,4){$=$}}
\put(4,22){\line(1,0){8}}
\put(12,20){\framebox(4,4){}}
\put(16,22){\line(1,0){8}}
\put(24,20){\framebox(4,4){$=$}}
\put(28,22){\line(1,0){8}}
\put(36,20){\framebox(4,4){}}
\put(40,22){\line(1,0){8}}
\put(48,20){\framebox(4,4){$=$}}
\put(52,22){\line(1,0){8}}
\put(60,20){\framebox(4,4){}}
\put(64,22){\line(1,0){8}}
\put(72,20){\framebox(4,4){$=$}}
\put(2,14){\line(0,1){6}}
\put(0,10){\framebox(4,4){}}
\put(2,10){\line(0,-1){6}}
\put(26,14){\line(0,1){6}}
\put(24,10){\framebox(4,4){}}
\put(26,10){\line(0,-1){6}}
\put(50,14){\line(0,1){6}}
\put(48,10){\framebox(4,4){}}
\put(50,10){\line(0,-1){6}}
\put(74,14){\line(0,1){6}}
\put(72,10){\framebox(4,4){}}
\put(74,10){\line(0,-1){6}}
\put(0,0){\framebox(4,4){$=$}}
\put(4,2){\line(1,0){8}}
\put(12,0){\framebox(4,4){}}
\put(16,2){\line(1,0){8}}
\put(24,0){\framebox(4,4){$=$}}
\put(28,2){\line(1,0){8}}
\put(36,0){\framebox(4,4){}}
\put(40,2){\line(1,0){8}}
\put(48,0){\framebox(4,4){$=$}}
\put(52,2){\line(1,0){8}}
\put(60,0){\framebox(4,4){}}
\put(64,2){\line(1,0){8}}
\put(72,0){\framebox(4,4){$=$}}
\end{picture}
}



Later on, in Section~\ref{sec:EstInfo}, we will consider noisy versions of such channels. 
As it turns out, the computation of the information 
rates of such channels also requires the computation 
of partition functions as in~(\ref{eqn:PartitionFunction}).

\section{Monte Carlo Methods for Partition Function Estimation}
\label{sec:BasicMonteCarlo}




One well-known method to estimate $\Gamma_{\!f} \eqdef 1/{Z_f}$
(and thus $Z_f$ itself) goes as follows.
\begin{trivlist}
\item\textbf{Ogata-Tanemura Method} \cite{OgTa:eip1981,PoGo:sapf1997}:
\begin{enumerate}
\item Draw samples $\x^{(1)}, \x^{(2)}, \ldots, \x^{(K)}$ from $\calS_f$ 
according to $p_f(\x)$ as in (\ref{eqn:Pdef}).
\item Compute
\begin{equation} \label{eqn:OgataTanemura}
\hat\Gamma_{\!f} = \frac{1}{K|\calS_f|} \sum_{k = 1}^K \frac{1}{f(\x^{(k)})}
\end{equation}
\end{enumerate}
\vspace{-2ex}
\hfill$\Box$
\end{trivlist}
It is easy to verify that $\E(\hat\Gamma_f) = 1/{Z_f}$. 

However, there are two major issues with this method.
First, there is the problem of generating the samples 
$\x^{(1)}, \x^{(2)}, \ldots, \x^{(K)}$
according to $p_f(\x)$. 
Ideally, we would like these samples to be independent, but (for the 
purposes of this paper) this is asking too much. In particular, we
will use Gibbs sampling~\cite{GG:srgd1984,MK:mct1998}, which
produces dependent samples. However, with naive Gibbs sampling,
the dependencies among the samples decay far too slowly for the 
estimate~(\ref{eqn:OgataTanemura}) to be useful for us (cf.\ the remarks 
in the Introduction).
We will see in Section~\ref{sec:TBGZ}, how this issue is eased by tree-based 
Gibbs sampling 
as proposed by Hamze and de~Freitas~\cite{FF:f2t2004c}.


Second, it is usually assumed 
that $f$ is strictly positive. In this case, $\calS_f = \calX$ 
and $|\calS_f| = |\calX|$ is known. 
However, this assumption excludes applications 
to constrained channels 
as in the example in Section~\ref{sec:PartitionFunction}.
Indeed, in that example,
we would have $f(\smpl{\x}{k})=1$ for all samples $\smpl{\x}{k}$,
and $|\calS_f|=Z_f$ is the desired unknown quantity.
We will address this issue in Section~\ref{subsec:ApplicationPFunction}.

With suitable modifications, which will address the mentioned issues,
the Ogata-Tanemura method will turn out to work well for 
the capacity of noiseless constrained 2-D channels.

However, for our second application---%
the information rate of noisy 2-D constrained source\andor{}channel models---%
the Ogata-Tanemura method turns out to be inadequate.
We will therefore resort to multilayer importance sampling 
as described below. 
We first describe the use of standard 
(single-layer) importance sampling 
to estimate $Z_f$.

\begin{trivlist}
\item\textbf{Importance Sampling} \cite{MK:mct1998,Neal:2001}:
\begin{enumerate}
\item Draw samples $\x^{(1)}, \x^{(2)}, \ldots, \x^{(K)}$ from $\calX$ 
according to some auxiliary probability distribution $q(\x) = \frac{1}{Z_g} g(\x)$,
where $g(\x)$ is a nonnegative function defined on $\calX$
satisfying $g(\x) \ne 0$ whenever $f(\x) \ne 0$.
\item Compute
\begin{equation} \label{eq:estImport}
\hat{R} = \frac{1}{K} \sum_{k=1}^K \frac{f(\x^{(k)})}{g(\x^{(k)})}
\end{equation}
\end{enumerate}
\vspace{-2ex}
\hfill$\Box$
\end{trivlist}
It is easy to verify that $\E(\hat{R}) = Z_f / Z_g$. 

The key issue with importance sampling is to find 
a useful function $g(\x)$ such that
\begin{itemize}
\item
$q(\x)$ is a good approximation of $p(\x)$ (so that $f(\x)/g(\x)$ 
does not wildly fluctuate),
\item
sampling from $q(\x)$ is feasible, and
\item
computing $Z_g$ is feasible.
\end{itemize}

An obvious choice for $g(\x)$ (and thus $q(\x)$) is 
\begin{equation}
g(\x) \eqdef f(\x)^{\alpha}
\end{equation}
for $0<\alpha<1$.
With this choice, any factorization of $f(\x)$ results in 
a factorization of $g(\x)$ that preserves the topology of the corresponding 
factor graph.
(Note, however, that this choice of $g(\x)$ is not helpful if $f(\x)$
is $\{0,1\}$-valued.)

In order to sample from $q(\x)$, 
we will again use tree-based Gibbs sampling (see Section~\ref{subsec:TreeBasedGibbsSampling}).

In a variation of the algorithm,
the estimator (\ref{eq:estImport}) of the ratio $Z_f / Z_g$ could be 
replaced by Bennett's acceptance ratio method~\cite{Bennett:76},
which is also known as bridge sampling \cite{MW:96}.

A function $g(\x)$ with all the required properties may be hard to find, 
or it may not exist. 
This problem is addressed by
multilayer importance sampling, which uses several auxiliary distributions.

%




\begin{figure}[t]
\setlength{\unitlength}{0.93mm}
\centering
\begin{picture}(82,84)(0,0)
\put(0,10){\drawgrid}
\thicklines
\put(26,42){\oval(24,84){}}
\put(72,42){\oval(20,84){}}
\put(2,79){\pos{c}{$A$}}
\put(26,79){\pos{c}{$B$}}
\put(50,79){\pos{c}{$A$}}
\put(74,79){\pos{c}{$B$}}
\end{picture}
\caption{\label{fig:2DGridPartition}%
Partition of \Fig{fig:2DGrid} into two cycle-free parts 
(one part inside the two ovals, the other part outside the ovals).}
\end{figure}
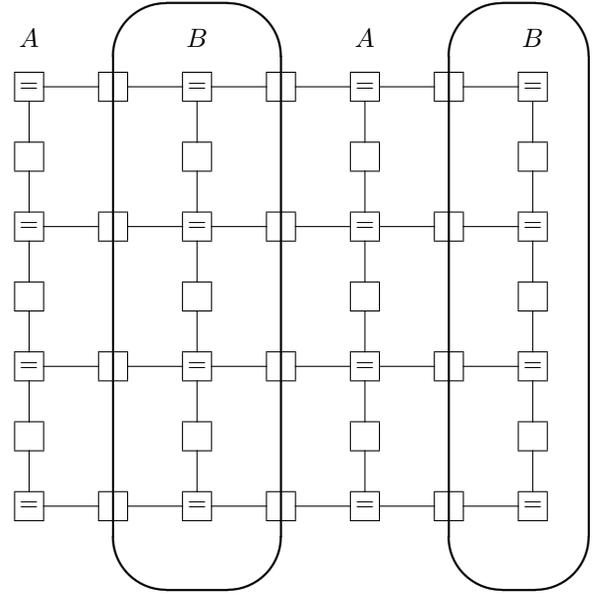


\begin{trivlist}
\item\textbf{Multilayer (Multi-Temperature) Importance Sampling:}

Multilayer importance sampling, as described here, 
is a variation of
annealed importance sampling as proposed by Neal~\cite{Neal:98}, 
\cite{Neal:2001}; see also~\cite{Jarz:97}.
We use $J$ parallel versions of importance sampling as follows. 
For $j=0,1\ldots, J$, let 
\begin{equation} \label{eqn:MultilayerG}
g_j(\x) \eqdef f(\x)^{\alpha_j}
\end{equation}
with $0 \le \alpha_J <  \cdots < \alpha_1 <  \alpha_0 =1$. 
Note that $Z_{g_0} = Z_f$ and
\begin{equation} \label{eqn:ratioZ}
\frac{Z_f}{Z_{g_J}} = \frac{Z_{g_0}}{Z_{g_1}} \frac{Z_{g_1}}{Z_{g_2}} \cdots \frac{Z_{g_{J-1}}}{Z_{g_J}}
\end{equation}
%
For $j=1, 2, \ldots, J$,
compute the ratio $Z_{g_{j-1}} / Z_{g_j}$
by importance sampling as described before:
\begin{enumerate}
\item
Draw samples $\x^{(1)}, \x^{(2)}, \ldots, \x^{(K)}$ from $q_j(\x) \propto g_j(\x)$.
\item
Compute
\begin{IEEEeqnarray}{rCl}
\hat R_j & = & \frac{1}{K} \sum_{k=1}^K \frac{g_{j-1}(\x^{(k)})}{g_j(\x^{(k)})}
               \IEEEeqnarraynumspace\\
& = & \frac{1}{K} \sum_{k=1}^K f(\x^{(k)})^{\alpha_{j-1}-\alpha_j}.
    \label{eqn:hatRj}
\vspace{-2ex}
\end{IEEEeqnarray}
\end{enumerate}
\hfill$\Box$
\end{trivlist}
Noting that $\E(\hat{R}_j) = Z_{g_{j-1}} / Z_{g_j}$,
we use 
$\prod_{j=1}^J \hat{R}_j$
as an estimate of $Z_f/Z_{g_J}$.


If the number of layers $J$ is large, 
$g_j(\x)$ is a good approximation of $g_{j-1}(\x)$ at each layer $j$.

It remains to find an estimate of $Z_{g_J}$, 
which tends to be easier than the original problem of estimating $Z_f$. 
In particular, for $\alpha_J=0$, we have 
$Z_{g_J} = |S_f|$, the cardinality of the support of $f$.
In our application (Section~\ref{sec:EstInfo}), it turns out that $Z_{g_J}$
can be computed efficiently by the tree-based Ogata-Tanemura method 
of Section~\ref{subsec:ApplicationPFunction}.


\section{Tree-Based Gibbs Sampling and Partition Function Estimation}
\label{sec:TBGZ}

\subsection{Tree-Based Gibbs Sampling}
\label{subsec:TreeBasedGibbsSampling}

Tree-based Gibbs sampling was proposed by Hamze and de Freitas~\cite{FF:f2t2004c}.
It works as follows.
Let $(A,B)$ be a partition of the index set $\{1, 2, \ldots,N\}$ such that,
\begin{itemize}
\item for fixed $\x_A$, the factor graph of $f(\x) = f(\x_A, \x_B)$ is cycle-free, and
\item for fixed $\x_B$, the factor graph of $f(\x) = f(\x_A, \x_B)$ is also cycle-free.
\end{itemize}
An example of such a partition is shown in \Fig{fig:2DGridPartition}.
Starting from some initial configuration 
$\smpl{\x}{0} = (\smpl{\x_A}{0}, \smpl{\x_B}{0})$, 
the samples
$\smpl{\x}{k} = (\smpl{\x_A}{k}, \smpl{\x_B}{k})$, $k=1,2,\ldots$, 
are created as follows:
first, $\smpl{\x_A}{k}$ is drawn from
\begin{equation}
p_f(\x_A \cond \x_B = \smpl{\x_B}{k-1}) \propto f(\x_A, \smpl{\x_B}{k-1});
\end{equation}
second, $\smpl{\x_B}{k}$ is drawn from
\begin{equation}
p_f(\x_B \cond \x_A = \smpl{\x_A}{k}) \propto f(\smpl{\x_A}{k}, \x_B).
\end{equation}

The point is that drawing these samples is easy 
(by means of backward-filtering forward-sampling, see the appendix) 
since the corresponding factor graphs are cycle-free.

As in standard Gibbs sampling, the samples 
$(\smpl{\x_A}{k}, \smpl{\x_B}{k})$ 
are eventually (i.e., for $k\rightarrow\infty$) drawn from $p_f$
(provided that the corresponding Markov chain is irreducible and aperiodic \cite{Bre:99}).
However, tree-based Gibbs sampling mixes faster than naive 
Gibbs sampling.

\subsection{Application to Partition Function Estimation}
\label{subsec:ApplicationPFunction}

With $A$ and $B$ as above, 
let
\begin{equation}
f_A(\x_A) \eqdef \sum_{\x_B} f(\x_A, \x_B),
\end{equation}
and
\begin{equation}
f_B(\x_B) \eqdef \sum_{\x_A} f(\x_A,\x_B).
\end{equation}
We then note that
\begin{IEEEeqnarray}{rCl}
Z_{f_A} & = & \sum_{\x_A} f_A(\x_A)
               \IEEEeqnarraynumspace\\
& = & Z_f,
\end{IEEEeqnarray}
i.e., $f_A$ (and analogously $f_B$) has the same partition function as $f$ itself.

We also note that samples 
$\smpl{\x_A}{1}$, $\smpl{\x_A}{2}$, \ldots, from
\begin{equation}
p_{f_A}(\x_A) \eqdef \frac{f_A(\x_A)}{Z_f}
= \sum_{\x_B} p_f(\x_A,\x_B) 
\end{equation}
can be obtained by tree-based Gibbs sampling 
as in Section~\ref{subsec:TreeBasedGibbsSampling}
simply by dropping the second component (the $B$-part) of 
the samples 
$(\smpl{\x_A}{1}, \smpl{\x_B}{1})$, $(\smpl{\x_A}{2}, \smpl{\x_B}{2})$, \ldots

We can now use the Ogata-Tanemura method (Section~\ref{sec:BasicMonteCarlo})
to estimate $\Gamma_f = 1/Z_f$ in two different ways. 
One way is to directly use the estimator (\ref{eqn:OgataTanemura}) 
with samples $\smpl{\x}{k}=(\smpl{\x_A}{k},\smpl{\x_B}{k})$
as in Section~\ref{subsec:TreeBasedGibbsSampling}. 
Another way is to apply the estimator (\ref{eqn:OgataTanemura}) to $f_A$, 
i.e., to compute
\begin{equation} \label{eqn:GammaA}
\hat\Gamma_{f_A} \eqdef \frac{1}{K |\calS_{f_A}|} \sum_{k=1}^K \frac{1}{f_A(\smpl{\x_A}{k})}
\end{equation}
which is also an estimate of $1/Z_f$.
The computation of
\begin{equation} \label{eqn:SampleMarginal}
f_A(\smpl{\x_A}{k}) = \sum_{\x_B} f(\smpl{\x_A}{k}, \x_B),
\end{equation}
which is required in (\ref{eqn:GammaA}), 
is easy since the corresponding factor graph is a tree;
in fact, the result of this computation is obtained for free 
as a by-product of tree-based Gibbs sampling as is explained in the appendix.
By symmetry, we also have an analogous estimate $\hat\Gamma_{f_B}$ defined as
\begin{equation} \label{eqn:GammaB}
\hat\Gamma_{f_B} \eqdef \frac{1}{K |\calS_{f_B}|} \sum_{k=1}^K \frac{1}{f_B(\smpl{\x_B}{k})}
\end{equation}

With the same samples 
$(\smpl{\x_A}{1}, \smpl{\x_B}{1})$, $(\smpl{\x_A}{2}, \smpl{\x_B}{2})$, \ldots,
estimating $1/Z_f$ from (\ref{eqn:GammaA}) and (\ref{eqn:GammaB}) 
tends to converge faster than (\ref{eqn:OgataTanemura}).
More importantly for this paper, 
the direct Ogata-Tanemura method (\ref{eqn:OgataTanemura})
cannot be used for constrained channels 
(cf.\ the example in Section~\ref{sec:PartitionFunction})
where $|\calS_f|=Z_f$ 
is the desired quantity. In contrast, 
the computation of $|\calS_{f_A}|$ in (\ref{eqn:GammaA})
and $|\calS_{f_B}|$ in (\ref{eqn:GammaB}) may be easy
in such cases as will be exemplified below.

\begin{figure}[t!]
\centering
\includegraphics[width=\linewidth]{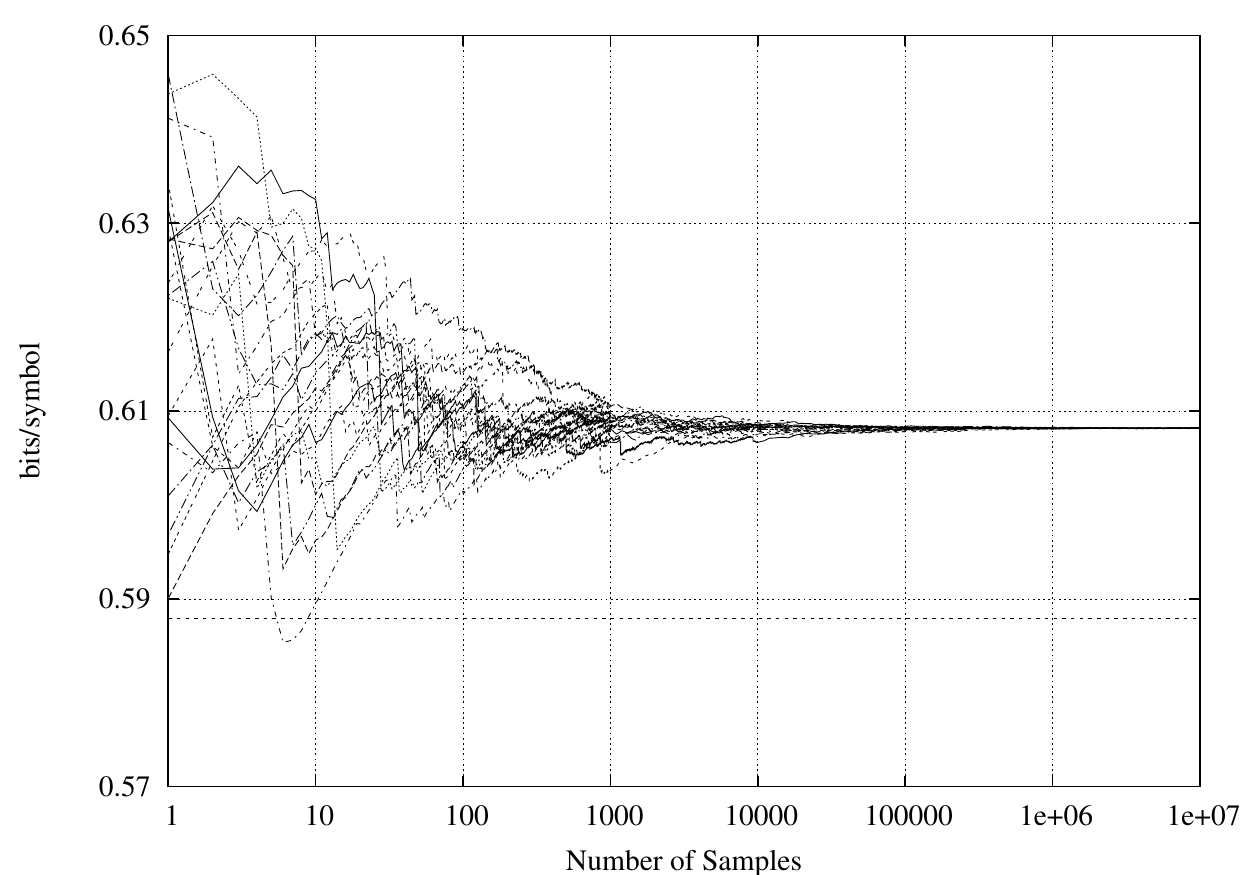}
\caption{\label{fig:p10}%
Estimated noiseless capacity (in bits/symbol) vs.\ the number of samples~$K$
for a $10\times 10$ grid with a $(1,\infty)$ constraint.
The plot shows 10 different sample paths, 
each with two estimates, one from $\Gamma_A$ and another from $\Gamma_B$.
The horizontal dotted line shows the infinite-size capacity for this constraint.}
\end{figure}

\section{Application to the Capacity of Noiseless 2-D Constrained Channels}
\label{sec:NoiselessCapNumEx}

We demonstrate the estimators (\ref{eqn:GammaA}) and (\ref{eqn:GammaB})
by their application to the example in Section~\ref{sec:PartitionFunction}, 
the 2-D $(1,\infty)$ runlength-limited constrained channel.

%

We will use factor graphs 
as in \Fig{fig:2DGrid}
with a partitioning as in \Fig{fig:2DGridPartition}.
In this example, 
the quantities $|\calS_{f_A}|$ and $|\calS_{f_B}|$, 
which are needed in (\ref{eqn:GammaA}) and (\ref{eqn:GammaB}), respectively,
are given by
\begin{eqnarray}
|\calS_{f_A}| & = & \sum_{\x_A} f(\x_A, \0) \\
|\calS_{f_B}| & = & \sum_{\x_B} f(\0, \x_B),
\end{eqnarray}
since
\begin{equation}
f(\x_A, \0) = \left\{ \begin{array}{ll}
  1, & \text{if $f_A(\x_A)>0$}\\
  0, & \text{if $f_A(\x_A)=0$},
\end{array}\right.
\end{equation}
and likewise for $f(\0, \x_B)$.
It follows that $|\calS_{f_A}|$ and $|\calS_{f_B}|$ are easy to compute
by sum-product message passing in 
the cycle-free factor graphs of $f(\x_A, \0)$ and $f(\0, \x_B)$, respectively.


\begin{figure}[t!]
\centering
\includegraphics[width=\linewidth]{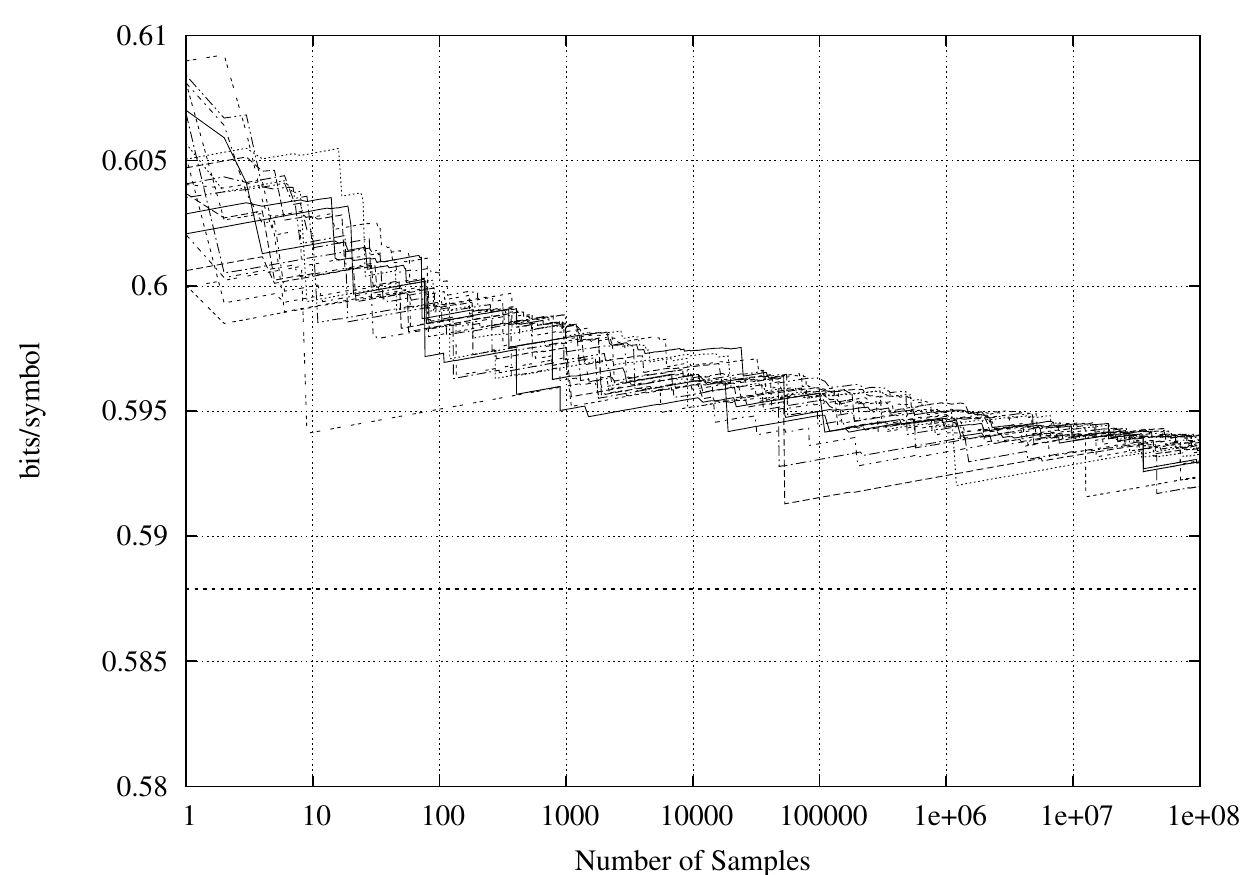}
\caption{\label{fig:p60-1}%
Estimated noiseless capacity (in bits/symbol) vs.\ the number of samples~$K$ 
for a $60\times 60$ grid with a $(1,\infty)$ constraint.
The plot shows 10 different sample paths, 
each with two estimates, one from $\Gamma_A$ and another from $\Gamma_B$.
The horizontal dotted line shows the infinite-size capacity for this constraint.}
\end{figure}

\begin{figure}[t!]
\includegraphics[width=\linewidth]{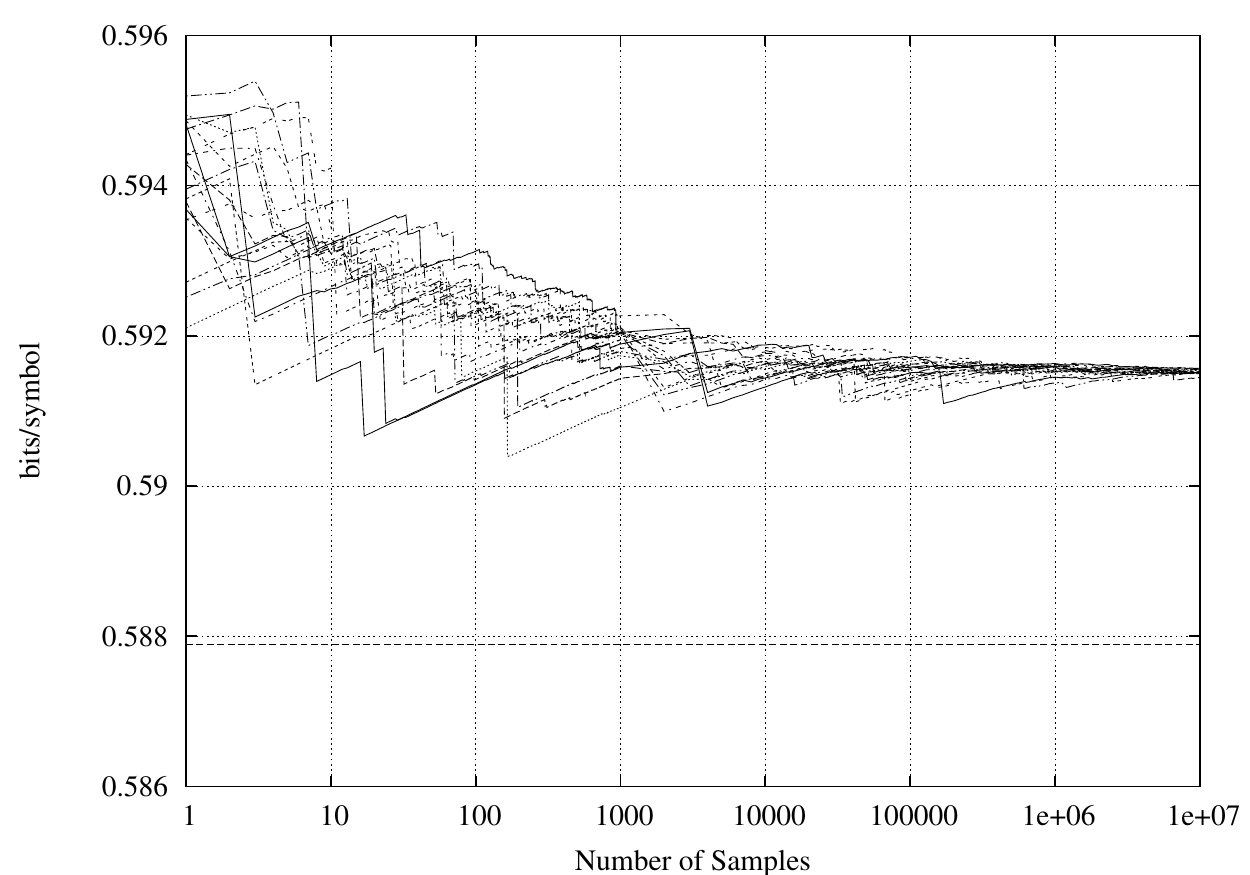}
\caption{\label{fig:p60-2}%
Same conditions as in Fig.~\ref{fig:p60-1}, but with strips of width two.}

\vspace{4.3mm}

\includegraphics[width=\linewidth]{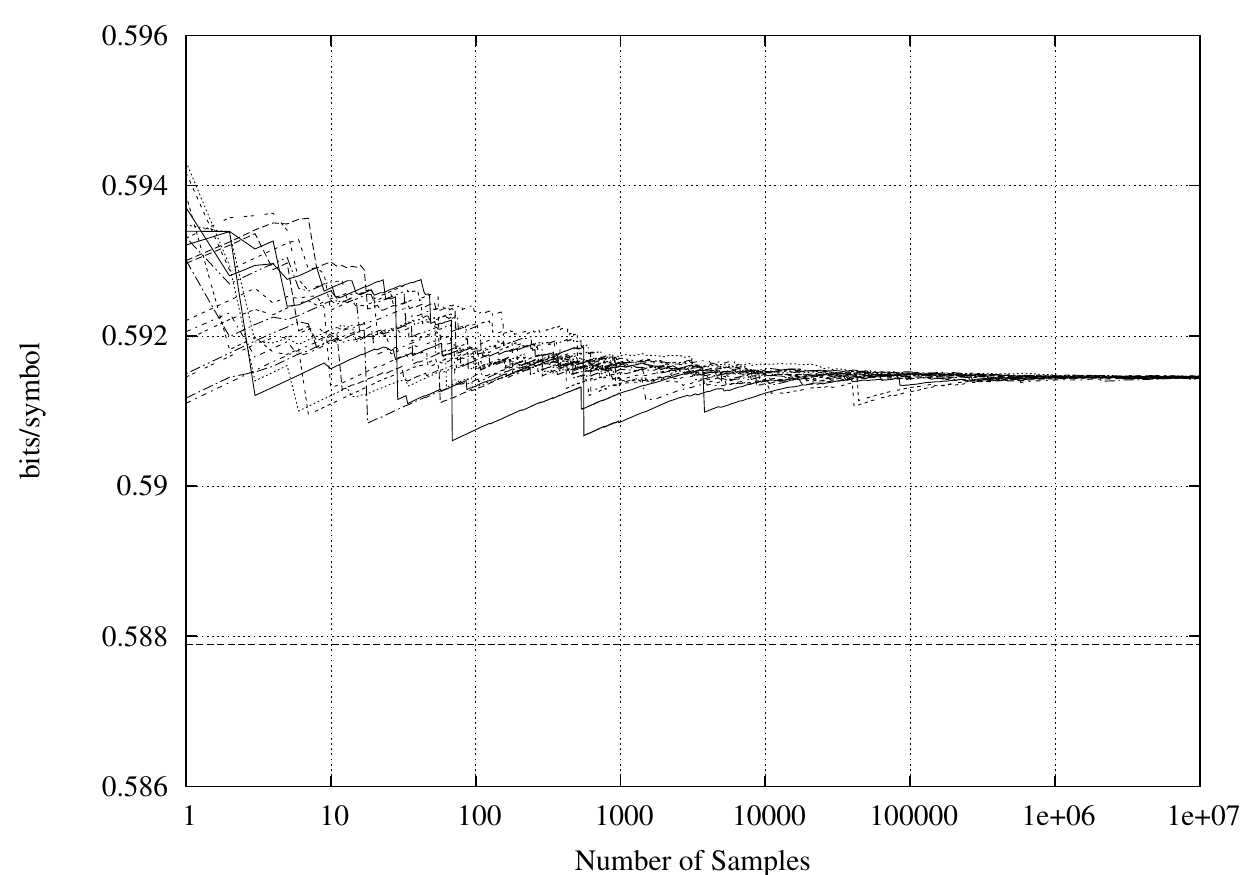}
\caption{\label{fig:p60-3}%
Same conditions as in Fig.~\ref{fig:p60-1}, but with strips of width three.}
\end{figure}

Some experimental results 
are shown in Figs. \ref{fig:p10} through~\ref{fig:p60-3}.
All figures refer to $f$ 
as in (\ref{eqn:NoAdjacentOnesFunction}) and (\ref{eqn:NoAdjacentOnesFactor})
and show the estimates of the finite-size capacity 
$C_M$ (\ref{eqn:CM})
obtained from (\ref{eqn:GammaA}) and (\ref{eqn:GammaB}) 
vs.~$K$ 
for several different experiments.

In \Fig{fig:p10}, we have $N=10\times 10$ and 
we obtain $C_{10}\approx 0.6082$ bits/symbol. 
In \Fig{fig:p60-1}, 
we have $N=60\times 60$, and there are issues with slow convergence.




In order to speed up the mixing 
and thus improving the convergence, 
we now partition the factor graph 
into vertical strips 
each containing $M\times 2$ or $M\times 3$ variables 
(rather than $M\times 1$ variables as in \Fig{fig:2DGridPartition}). 
Exact sum-product message passing is still possible on such strips, 
e.g., by converting the strip into an equivalent cycle-free factor graph. 
The computation time is exponential in the width of the strip,
but the faster mixing 
results in a substantial reduction of total computation time
for strips of moderate width.

Simulation results for strips of width 2 and 3
are shown in Figs. \ref{fig:p60-2} and~\ref{fig:p60-3},
respectively, both for a grid of size $N=60\times 60$.
Note that the convergence is much better than in \Fig{fig:p60-1},
and we obtain $C_{60} \approx 0.5914$ bits/symbol.


Also shown in these figures, as a horizontal dotted line, is the 
infinite-size capacity $C_\infty \approx\ 0.5879$ bits/symbol 
from \cite{NZ:2000},
which (in this example) is a lower bound on the finite-size capacity.

\section{Estimating the Information Rate of Noisy 2-D Source\andor{}Channel Models}
\label{sec:EstInfo}

\subsection{The Problem}

We now consider the problem of computing the information rate 
of noisy 2-D source\andor{}channel models. 
Although the focus of this paper is on constrained channels,
the proposed approach can also be applied to other 2-D source\andor{}channel models
such as 2-D channels with intersymbol interference.

For a 2-D grid of size $N=M\times M$ (as before),
let $\X = \{X_1,X_2, \ldots, X_N\}$ be the source process
(i.e., the input process of the channel) 
and let
$\Y = \{Y_1,Y_2, \ldots, Y_N\}$ be the output process of the channel;
$\X$ takes values in $\calX$ (as defined in Section~\ref{sec:PartitionFunction})
and $\Y$ takes values in $\R^N$.
We wish to compute 
the mutual information rate
\begin{equation} \label{eqn:IR}
\frac{1}{N}I(\X;\Y) = \frac{1}{N}\big(H(\Y) - H(\Y \cond \X)\big)
\end{equation}
for cases where $p(\x,\y)$ has a suitable factor graph. 
In particular, 
we will focus on the case where the channel is memoryless, i.e.,
\begin{equation} \label{eqn:MemorylessChannel}
p(\y \cond \x) = \prod_{n=1}^N p(y_n \cond x_n),
\end{equation}
and where the channel input distribution
$p(\x)$ has a factorization with a factor graph as 
in \Fig{fig:2DGrid}. 
It then follows that $p(\x,\y)$ has a factor graph as in 
\Fig{fig:SourceChannelGraph}.

In many cases of practical interest, $H(\Y \cond \X)$ is analytically available,
see our numerical experiments in Section~\ref{sec:NumInfo}.
In such cases, the problem of computing 
the mutual information rate (\ref{eqn:IR}) reduces 
to computing 
\begin{equation} \label{eqn:HY}
H(\Y) = \E\big[ \mbox{$-\log_2 p(\Y)$} \big].
\end{equation}

If $H(\Y \cond \X)$  is not analytically available, it can be computed
by the same method as $H(\Y)$,
see~\cite[Section~III]{ALVKZ:sbcir2006}.

\begin{figure}[t]
\setlength{\unitlength}{0.94mm}
\centering
\begin{picture}(87,73)(0,-9)
\small
%
\put(0,60){\framebox(4,4){$=$}}
 \put(4,60){\line(4,-3){4}}
 \put(8,54){\framebox(3,3){}}
 \put(11,54){\line(4,-3){4}}     \put(14,53){$Y_1$}
\put(4,62){\line(1,0){8}}        \put(8,63){\pos{bc}{$X_1$}}
\put(12,60){\framebox(4,4){}}
\put(16,62){\line(1,0){8}}
\put(24,60){\framebox(4,4){$=$}}
 \put(28,60){\line(4,-3){4}}
 \put(32,54){\framebox(3,3){}}
 \put(35,54){\line(4,-3){4}}     \put(38,53){$Y_2$}
\put(28,62){\line(1,0){8}}       \put(32,63){\pos{bc}{$X_2$}}
\put(36,60){\framebox(4,4){}}
\put(40,62){\line(1,0){8}}
\put(48,60){\framebox(4,4){$=$}}
 \put(52,60){\line(4,-3){4}}
 \put(56,54){\framebox(3,3){}}
 \put(59,54){\line(4,-3){4}}     \put(62,53){$Y_3$}
\put(52,62){\line(1,0){8}}       \put(56,63){\pos{bc}{$X_3$}}
\put(60,60){\framebox(4,4){}}
\put(64,62){\line(1,0){8}}
\put(72,60){\framebox(4,4){$=$}}
 \put(76,60){\line(4,-3){4}}
 \put(80,54){\framebox(3,3){}}
 \put(83,54){\line(4,-3){4}}
\put(2,54){\line(0,1){6}}
\put(0,50){\framebox(4,4){}}
\put(2,50){\line(0,-1){6}}
\put(26,54){\line(0,1){6}}
\put(24,50){\framebox(4,4){}}
\put(26,50){\line(0,-1){6}}
\put(50,54){\line(0,1){6}}
\put(48,50){\framebox(4,4){}}
\put(50,50){\line(0,-1){6}}
\put(74,54){\line(0,1){6}}
\put(72,50){\framebox(4,4){}}
\put(74,50){\line(0,-1){6}}
\put(0,40){\framebox(4,4){$=$}}
 \put(4,40){\line(4,-3){4}}
 \put(8,34){\framebox(3,3){}}
 \put(11,34){\line(4,-3){4}}
\put(4,42){\line(1,0){8}}
\put(12,40){\framebox(4,4){}}
\put(16,42){\line(1,0){8}}
\put(24,40){\framebox(4,4){$=$}}
 \put(28,40){\line(4,-3){4}}
 \put(32,34){\framebox(3,3){}}
 \put(35,34){\line(4,-3){4}}
\put(28,42){\line(1,0){8}}
\put(36,40){\framebox(4,4){}}
\put(40,42){\line(1,0){8}}
\put(48,40){\framebox(4,4){$=$}}
 \put(52,40){\line(4,-3){4}}
 \put(56,34){\framebox(3,3){}}
 \put(59,34){\line(4,-3){4}}
\put(52,42){\line(1,0){8}}
\put(60,40){\framebox(4,4){}}
\put(64,42){\line(1,0){8}}
\put(72,40){\framebox(4,4){$=$}}
 \put(76,40){\line(4,-3){4}}
 \put(80,34){\framebox(3,3){}}
 \put(83,34){\line(4,-3){4}}
\put(2,34){\line(0,1){6}}
\put(0,30){\framebox(4,4){}}
\put(2,30){\line(0,-1){6}}
\put(26,34){\line(0,1){6}}
\put(24,30){\framebox(4,4){}}
\put(26,30){\line(0,-1){6}}
\put(50,34){\line(0,1){6}}
\put(48,30){\framebox(4,4){}}
\put(50,30){\line(0,-1){6}}
\put(74,34){\line(0,1){6}}
\put(72,30){\framebox(4,4){}}
\put(74,30){\line(0,-1){6}}
\put(0,20){\framebox(4,4){$=$}}
 \put(4,20){\line(4,-3){4}}
 \put(8,14){\framebox(3,3){}}
 \put(11,14){\line(4,-3){4}}
\put(4,22){\line(1,0){8}}
\put(12,20){\framebox(4,4){}}
\put(16,22){\line(1,0){8}}
\put(24,20){\framebox(4,4){$=$}}
 \put(28,20){\line(4,-3){4}}
 \put(32,14){\framebox(3,3){}}
 \put(35,14){\line(4,-3){4}}
\put(28,22){\line(1,0){8}}
\put(36,20){\framebox(4,4){}}
\put(40,22){\line(1,0){8}}
\put(48,20){\framebox(4,4){$=$}}
 \put(52,20){\line(4,-3){4}}
 \put(56,14){\framebox(3,3){}}
 \put(59,14){\line(4,-3){4}}
\put(52,22){\line(1,0){8}}
\put(60,20){\framebox(4,4){}}
\put(64,22){\line(1,0){8}}
\put(72,20){\framebox(4,4){$=$}}
 \put(76,20){\line(4,-3){4}}
 \put(80,14){\framebox(3,3){}}
 \put(83,14){\line(4,-3){4}}
\put(2,14){\line(0,1){6}}
\put(0,10){\framebox(4,4){}}
\put(2,10){\line(0,-1){6}}
\put(26,14){\line(0,1){6}}
\put(24,10){\framebox(4,4){}}
\put(26,10){\line(0,-1){6}}
\put(50,14){\line(0,1){6}}
\put(48,10){\framebox(4,4){}}
\put(50,10){\line(0,-1){6}}
\put(74,14){\line(0,1){6}}
\put(72,10){\framebox(4,4){}}
\put(74,10){\line(0,-1){6}}
\put(0,0){\framebox(4,4){$=$}}
 \put(4,0){\line(4,-3){4}}
 \put(8,-6){\framebox(3,3){}}
 \put(11,-6){\line(4,-3){4}}
\put(4,2){\line(1,0){8}}
\put(12,0){\framebox(4,4){}}
\put(16,2){\line(1,0){8}}
\put(24,0){\framebox(4,4){$=$}}
 \put(28,0){\line(4,-3){4}}
 \put(32,-6){\framebox(3,3){}}
 \put(35,-6){\line(4,-3){4}}
\put(28,2){\line(1,0){8}}
\put(36,0){\framebox(4,4){}}
\put(40,2){\line(1,0){8}}
\put(48,0){\framebox(4,4){$=$}}
 \put(52,0){\line(4,-3){4}}
 \put(56,-6){\framebox(3,3){}}
 \put(59,-6){\line(4,-3){4}}
\put(52,2){\line(1,0){8}}
\put(60,0){\framebox(4,4){}}
\put(64,2){\line(1,0){8}}
\put(72,0){\framebox(4,4){$=$}}
 \put(76,0){\line(4,-3){4}}
 \put(80,-6){\framebox(3,3){}}
 \put(83,-6){\line(4,-3){4}}
\end{picture}
\caption{\label{fig:SourceChannelGraph}%
Extension of \Fig{fig:2DGrid} to a
factor graph of $p(\x) p(\y\cond \x)$
with $p(\y \cond \x)$ as in~(\ref{eqn:MemorylessChannel}).
}
\end{figure}

\subsection{The Method}
\label{sec:EstInfoMethod}

As in~\cite{ALVKZ:sbcir2006}, 
we approximate the expectation 
in (\ref{eqn:HY}) by the empirical average
\begin{equation} \label{eqn:HYE}
H(\Y) \approx -\frac{1}{L}\sum_{\ell = 1}^{L} \log_2\!\big(p(\y^{(\ell)})\big),
\end{equation}
where samples
$\y^{(1)},\y^{(2)}, \ldots, \y^{(L)}$
are drawn according to $p(\y)$.
The problem of estimating the mutual information rate is
thus reduced to 
\begin{itemize}
\item
creating samples $\y^{(\ell)}$ and 
\item
computing $p(\y^{(\ell)})$ for each sample.
\end{itemize}

If $p(\x,\y)$ has a cycle-free factor graph 
(and if $|\calX_1|, |\calX_2|, \ldots, |\calX_N|$ are not too large),
then both tasks can be carried out in a single-loop algorithm 
as in \cite{ALVKZ:sbcir2006}. 
In this paper, however, we assume that no such factor graph exists
and we propose a double-loop algorithm 
(with an outer loop and an inner loop)
to carry out these tasks.
In the outer loop, 
we create samples $\y^{(1)}, \ldots, \y^{(L)}$ as follows. 
\begin{enumerate}
\item
Draw samples $\x^{(1)}, \ldots, \x^{(L)}$ according to $p(\x)$. 
In simple cases (such as unconstrained channels with i.i.d.\ input),
this may be trivial; in general, however, we do this
by tree-based Gibbs sampling (as in Section~\ref{subsec:TreeBasedGibbsSampling})
using the factor graph of $p(\x)$.
\item
For $\ell=1,\ldots,L$, draw $\smpl{\y}{\ell}$
from $p_{\Y|\X}(\y \cond \smpl{\x}{\ell})$, 
i.e., by simulating the channel with input $\smpl{\x}{\ell}$.
\end{enumerate}
%
In the inner loop, 
we compute an estimate of
\begin{equation}  \label{eqn:PY}
p(\y^{(\ell)}) = \sum_{\x \in\calX} p(\vct{x})\, p_{\Y|\X}(\smpl{\y}{\ell} \cond \vct{x})
\end{equation}
as follows. 
Note that, for fixed $\ell$, 
the right-hand side of (\ref{eqn:PY}) is the partition function 
$Z_{f_\ell}$ of
\begin{equation} \label{eqn:fell}
f_\ell(\x) \eqdef p(\vct{x})\, p_{\Y|\X}(\smpl{\y}{\ell} \cond \vct{x}), 
\end{equation}
which has a suitable factor graph (as, e.g., in \Fig{fig:SourceChannelGraph}).
In principle, 
we can thus estimate (\ref{eqn:PY})
by any of the methods of Section~\ref{sec:BasicMonteCarlo}. 
It turns out, however, that 
only multilayer importance sampling 
is able to handle the more demanding cases 
(as will be explained in our numerical experiments in Section~\ref{sec:NumInfo})
while
the other methods of Section~\ref{sec:BasicMonteCarlo}
suffer from slow and erratic convergence. 


%

\begin{figure}[t]
\centering
\includegraphics[width= \linewidth]{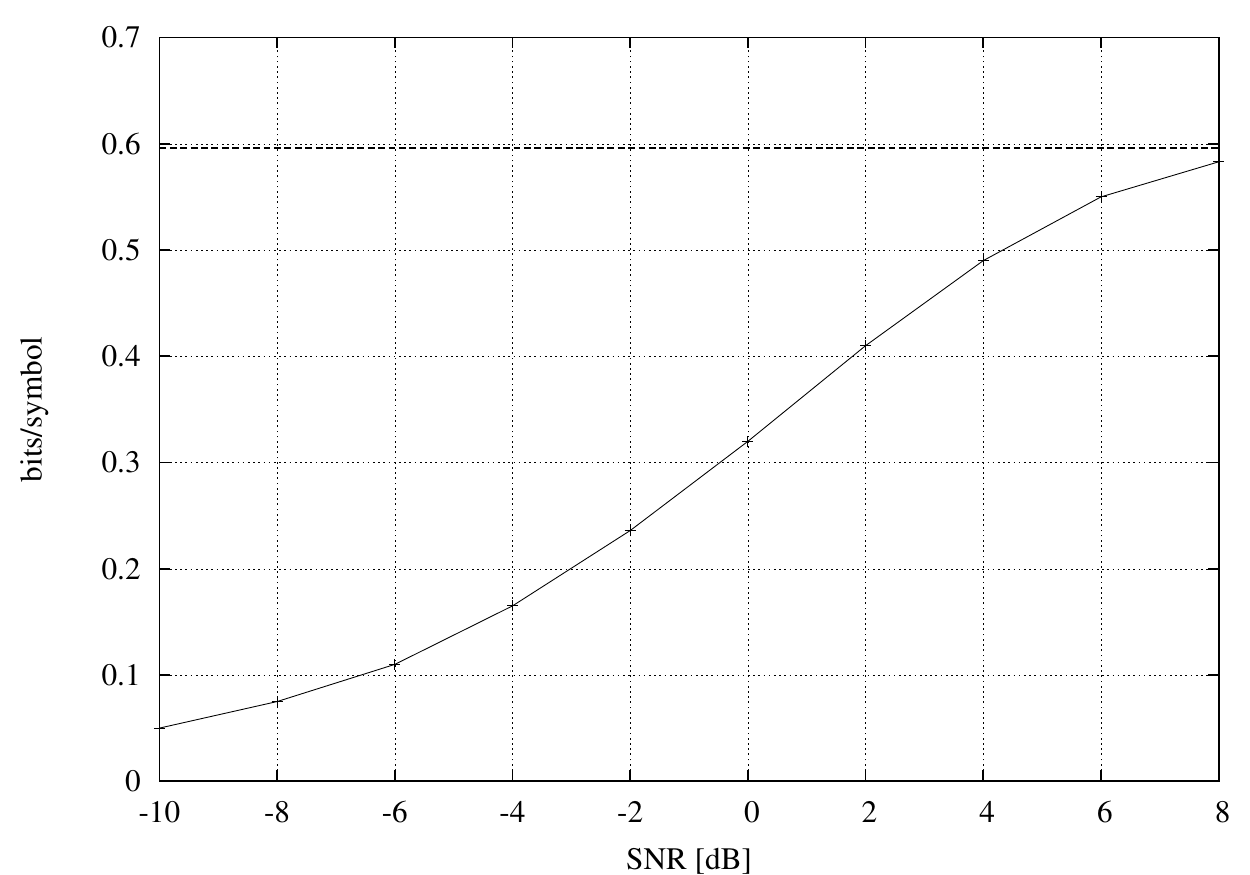}
\caption{\label{fig:SNR2}%
Uniform-input information rate (in bits/symbol) vs.\ SNR 
for a $24\times 24$ channel with a $(1,\infty)$ constraint and additive
white Gaussian noise. The horizontal dotted line shows the 
noiseless capacity of this channel.}
\end{figure}

\subsection{Numerical Experiments}
\label{sec:NumInfo}

In our numerical experiments, we consider a noisy 
version of the example in Section~\ref{sec:PartitionFunction},
i.e., a noisy version of the \mbox{2-D} $(1,\infty)$ runlength-limited constrained channel.
We assume that the channel input distribution $p(\x)$ is uniform over the 
allowed configurations, i.e., 
$p(\x) = p_f(\x)$ with $f$ as in (\ref{eqn:NoAdjacentOnesFunction}),
and we assume that the noise is additive white Gaussian (and independent of $\X$),
i.e., $p(\y\cond \x)$ is a product as in (\ref{eqn:MemorylessChannel}) with factors
\begin{equation} \label{eqn:NoisyChanModNumEx}
p(y_n \cond x_n) = \frac{1}{\sqrt{2\pi\sigma^2}}
          \exp\!\bigg(-\frac{1}{2\sigma^2} 
          \Big(y_n - (-1)^{x_n}\Big)^2 \bigg)
\end{equation}
and thus
\begin{equation}
H(\Y \cond \X) =  \frac{N}{2}\log_2(2\pi e\sigma^2).
\end{equation}
We will use the signal-to-noise ratio (SNR) 
defined as 
\begin{equation}
\text{SNR} \eqdef \frac{1}{\sigma^2},
\end{equation}
which we will specify in dB, i.e., $10\log_{10}(\text{SNR})$.


\begin{figure}
\centering
\includegraphics[width = \linewidth]{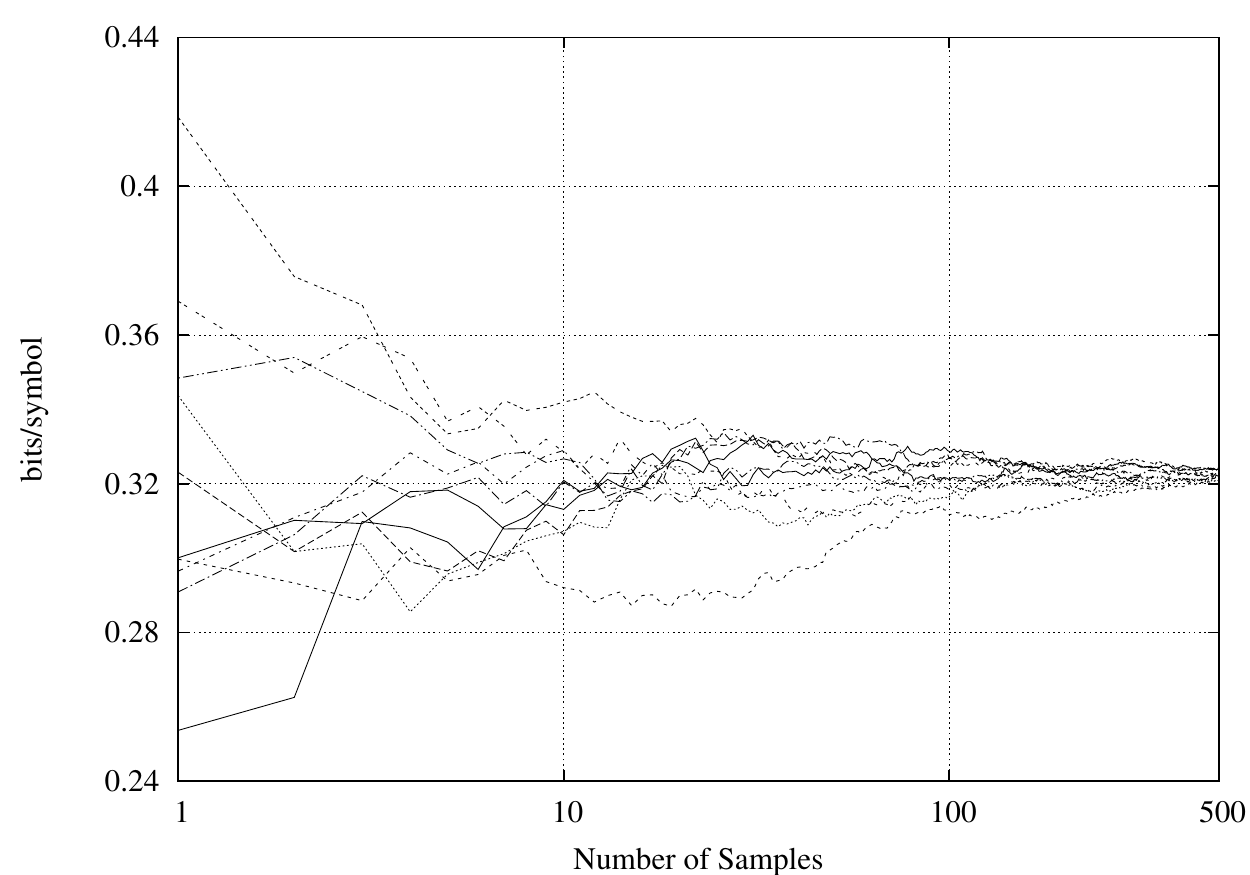}
\caption{\label{fig:IR_zero}%
Estimated information rate (in bits/symbol) vs.\ the number of samples $L$
for a noisy $24\times 24$ $(1,\infty)$ constraint at 0 dB.
The plot shows $12$ different sample paths.}
\vspace{5.1mm}
\centering
\includegraphics[width = \linewidth]{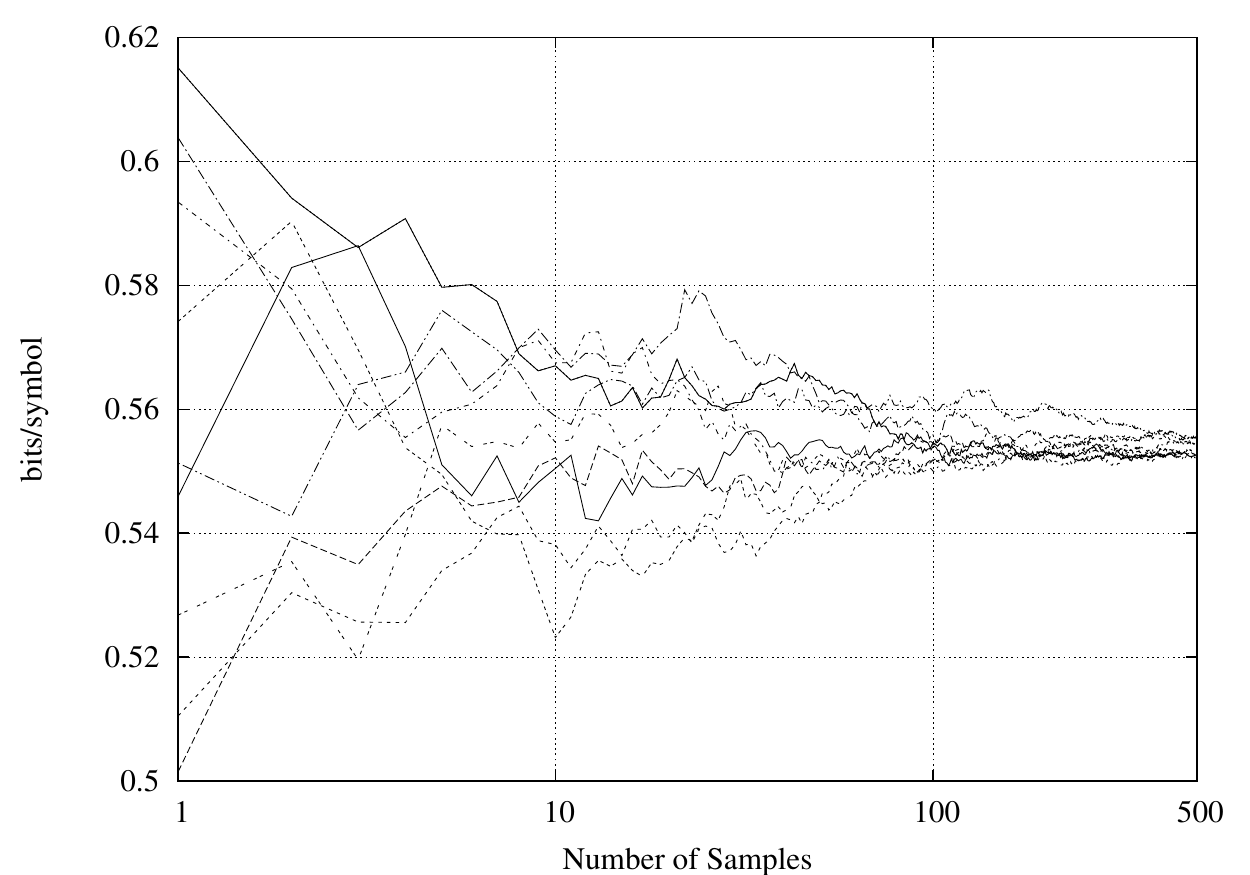}
\caption{\label{fig:IR_six}%
Estimated information rate (in bits/symbol) vs.\ the number of samples $L$
for a noisy $24\times 24$ $(1,\infty)$ constraint at 6 dB.
The plot shows $12$ different sample paths.}
\end{figure}

The grid size in all the plots is $N=24\times 24$
and
the parameters $\alpha_j$ in (\ref{eqn:MultilayerG})
are set to $\alpha_j = 2^{-j}$, for $j=1,2,\ldots, J$.

\Fig{fig:SNR2} shows the computed information rate 
vs.\ the SNR 
over the interval $[-10,8]$~dB. 
The horizontal dotted line in~\Fig{fig:SNR2} shows the 
capacity of the noiseless version of this channel, 
which is about $0.596$ bits per symbol. 

Figs. \ref{fig:IR_zero} and~\ref{fig:IR_six}
illustrate the convergence of the outer loop 
of the proposed double-loop algorithm
at 0~dB and at 6~dB, respectively.
Both figures show the estimated information rate
vs.\ the number of samples $L$ in (\ref{eqn:HYE})
for 12 different sample paths.

As for the inner loop, 
the required number of layers $J$ in (\ref{eqn:ratioZ}) 
depends on the SNR. As the SNR increases 
(or equivalently as $\sigma^2$ decreases),
the function $f_\ell(\x)$ in (\ref{eqn:fell})
tends to have more isolated modes. Therefore, in order to obtain
a good approximation at each level of multilayer importance 
sampling, larger values of $J$ are required for higher SNR.
In our numerical experiments at 0~dB
and 6~dB,
$J$ is set to 3 and 6,
respectively.

\begin{figure}
\centering
\includegraphics[width = \linewidth]{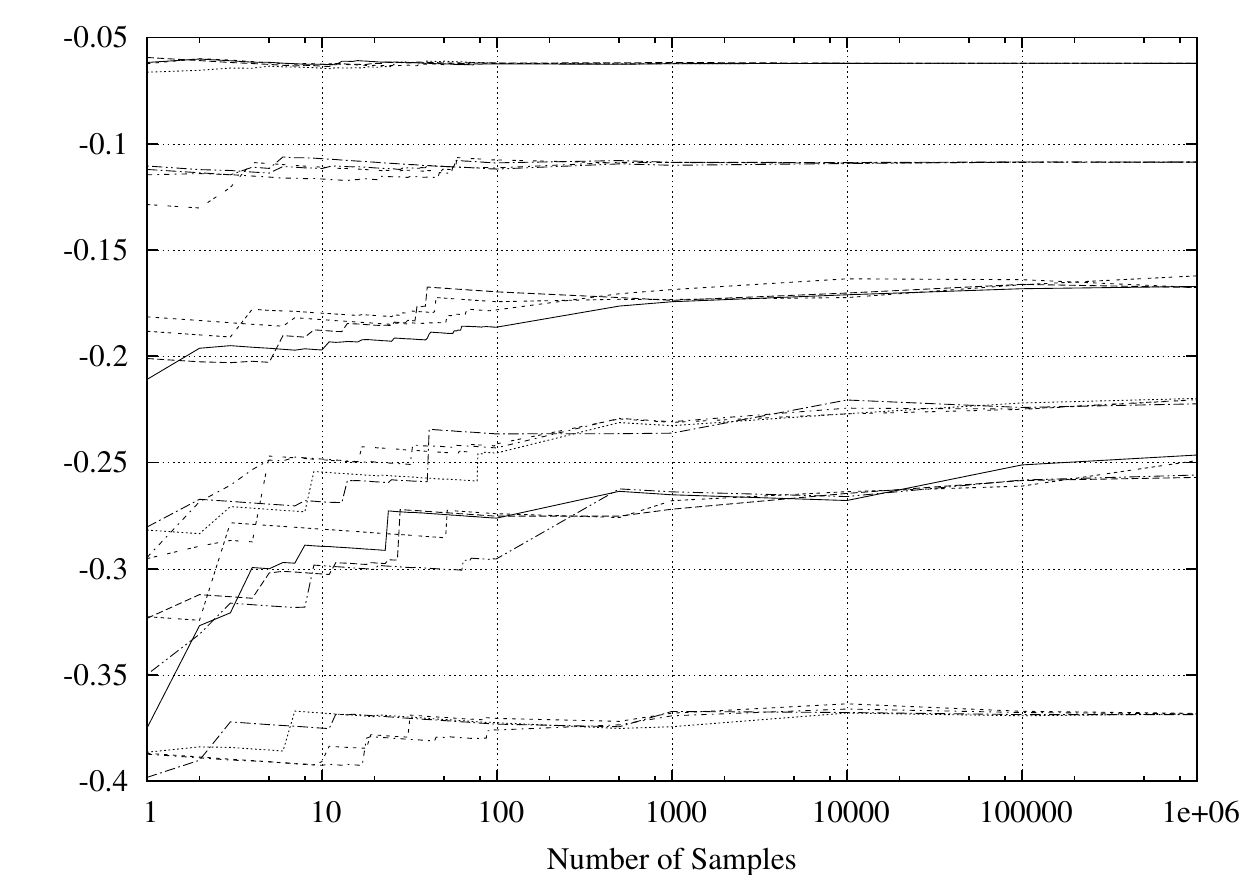}
\caption{\label{fig:Lev1}%
Computed $\log_2 \hat R_j$ as in (\ref{eqn:hatRj}), for $j=1, 2, \ldots,6$ vs.\ the number 
of samples $K$ for a noisy $24\times 24$ $(1,\infty)$ runlength-limited 
constraint at $6$ dB. The plot shows $\log_2 \hat R_6, \log_2 \hat R_5, \ldots, \log_2 \hat R_1$ from top to bottom.}
\vspace{5mm}
\includegraphics[width = \linewidth]{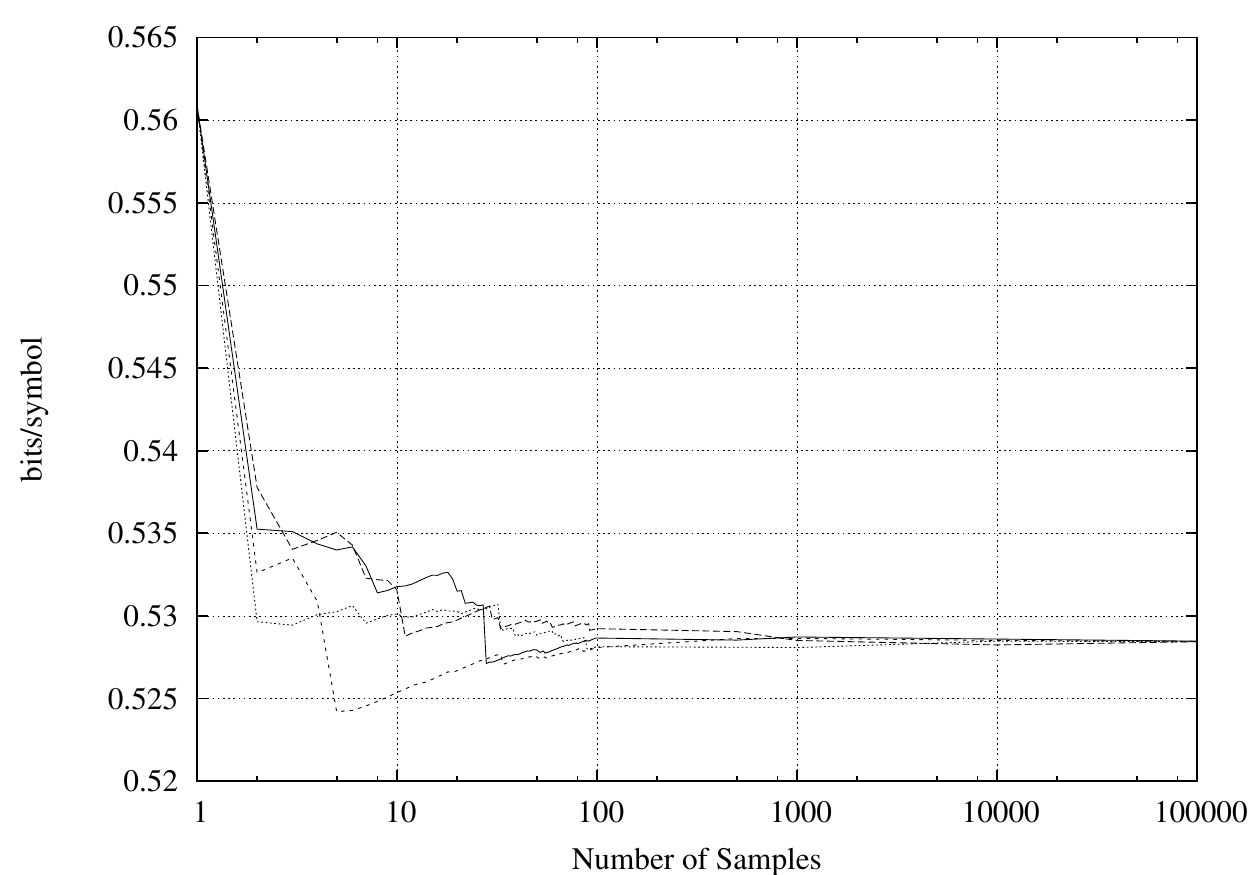}
\caption{\label{fig:Lev2}%
Estimated $\log_2(Z_{g_6})/N$ vs.\ the number of samples $K$
for a noisy $24\times 24$ $(1,\infty)$ runlength-limited constraint at $6$ dB.}
\end{figure}

\Fig{fig:Lev1} shows the
convergence of $\log_2 \hat R_j$ as in (\ref{eqn:hatRj})
for $j=1,2,\ldots,6$, for a fixed output sample at 6~dB.

The value of $Z_{g_J}$ is estimated by the 
tree-based Ogata-Tanemura method of Section~\ref{subsec:ApplicationPFunction}.
\Fig{fig:Lev2} shows the convergence of the estimate of
$\log_2 (Z_{g_6})/N$ according to (\ref{eqn:GammaA}) 
for four different sample paths.

\subsection{Remarks}

In statistical physics, 
the partition function typically has the form
\begin{equation}
Z =  \sum_{\x \in \calX} e^{-E(\x)/T},
\end{equation}
where $T$ is the temperature 
and $E(\x)$ is the energy of the configuration $\x$.
We therefore point out that the noise variance $\sigma^2$ in (\ref{eqn:NoisyChanModNumEx}) 
may be viewed as the temperature parameter of 
the partition function $Z_{f_\ell}$ of (\ref{eqn:fell}). 
It is well known in statistical physics that 
computing the partition function is harder at low temperatures than at high temperatures. 
Similarly, we observe that computing 
the partition function $Z_{f_\ell}$ of (\ref{eqn:fell})
is harder at high SNR than at low SNR;
in particular, at high SNR, 
more layers (higher values of $J$)
are required for multilayer (multi-temperature) importance sampling.

We also note that, in the examples of Section~\ref{sec:NumInfo},
the choice of the parameters $\alpha_j = 2^{-j}$ in (\ref{eqn:MultilayerG})
is somewhat arbitrary. 
It is possible that other choices of these parameters result in faster convergence.


%


\section{Conclusion}
\label{sec:Conclusion}

Monte Carlo methods have been highly succesful
in computing the information rate of source\andor{}channel models 
with \mbox{1-D} memory. 
The extension of such methods to source\andor{}channel models with 2-D memory 
has been an open research problem. 
In this paper, we develop such methods 
with a focus on the (difficult) case of channels with input constraints,
with or without noise.
In contrast to previous techniques,
which either use generalized belief propagation
or compute only bounds on the information rate, 
the Monte Carlo algorithms of this paper are 
guaranteed to converge (asymptotically) to the desired information rate.
A key role in the proposed algorithms is played by tree-based Gibbs sampling by 
Hamze and de Freitas, which we have shown to yield an estimate of the partition
function as a by-product.
The success of the proposed methods is exemplified 
by \Fig{fig:SNR2}, which (to the best of our knowledge) 
is the first such plot for a noisy 2-D channel. 
We also note that the extension of the proposed methods to computing upper 
and lower bounds on the information rate as in~\cite[Section~VI]{ALVKZ:sbcir2006} 
is straightforward.


\section*{Acknowledgement}

The authors wish to thank Radford Neal for 
helpful discussions and advice on annealed importance sampling.
The authors also wish to thank David MacKay and Iain Murray 
for pointing out to us \cite{FF:f2t2004c}, and the reviewers and the Associate Editor 
for helpful comments.

\appendix[Sampling from Markov Chains]
\label{sec:SMC}

We recall some pertinent facts about the simulation of Markov chains 
and cycle-free factor graphs.
Let $p(\x) = p(x_1,\ldots,x_n)$ be the probability mass function 
of a Markov chain. 
If $p(\x)$ is given in the form
\begin{equation} \label{eqn:PxChainRule}
p(\x) = p(x_1) \prod_{k=2}^n p(x_k|x_{k-1}),
\end{equation}
then it is obvious how to draw i.i.d.\ samples according to $p(\x)$. 
Now consider the case where $p(\x)$ is not given in the form (\ref{eqn:PxChainRule}),
but in the more general form
\begin{equation} \label{eqn:PxCostChain}
p(\x) \propto \prod_{k=2}^n g_k(x_{k-1},x_k)
\end{equation}
with general factors $g_k$.
It is then still easy to draw i.i.d.\ samples according to $p(\x)$,
which may be seen as follows. 
First, a probability mass function of the form (\ref{eqn:PxCostChain})
can be rewritten in the form (\ref{eqn:PxChainRule}) 
(which allows efficient simulation).
Second, this reparameterization of $p(\x)$ 
may be efficiently carried out by backward
sum-product message passing, as will be detailed below. 
The resulting algorithm is known as ``backward-filtering forward-sampling'' 
(or, in a time-reversed version, as ``forward-filtering backward-sampling'') 
\cite{GaLo:mcmc2006}.

Specifically, let $\msgb{\mu}{X_k}$ be the backward sum-product message 
along the edge $X_k$ in the factor graph 
of~(\ref{eqn:PxCostChain}), as is illustrated in Fig.~\ref{fig:FFGPxCostChainMsgb}
(cf.\ \cite{Lg:ifg2004}).
We then have 
$\msgb{\mu}{X_n}(x_n)=1$ and 
\begin{IEEEeqnarray}{rCl}
\msgb{\mu}{X_k}(x_k) 
& \eqdef &  \sum_{x_{k+1}} g_{k+1}(x_k,x_{k+1}) \msgb{\mu}{X_{k+1}}(x_{k+1}) 
  \label{eqn:SumProdMsgbXk}\\
& = & \sum_{x_{k+1},\ldots,x_n}\, \prod_{m=k+1}^n g_m(x_{m-1},x_m)
\end{IEEEeqnarray}
for $k=n-1, n-2, \ldots, 1$.
Then
\begin{IEEEeqnarray}{rCl}
p(x_1)  & = & \sum_{x_2,\ldots,x_n} p(x_1,\ldots,x_n) \\
& \propto & \msgb{\mu}{X_1}(x_1)
\end{IEEEeqnarray}
and
\begin{equation} \label{eqn:transprobforw}
p(x_k|x_{k-1}) = \frac{g_k(x_{k-1},x_k) \msgb{\mu}{X_k}(x_k)}{\msgb{\mu}{X_{k-1}}(x_{k-1})}
\end{equation}
for $k=2,\ldots,n$. 
The proof of (\ref{eqn:transprobforw}) follows from noting that
\begin{equation}
p(x_{k-1}) = \gamma \msgf{\mu}{X_{k-1}}(x_{k-1}) \msgb{\mu}{X_{k-1}}(x_{k-1})
\end{equation}
and
\begin{equation}
p(x_{k-1},x_k) = \gamma \msgf{\mu}{X_{k-1}}(x_{k-1}) g_k(x_{k-1},x_k) \msgb{\mu}{X_k}(x_k)
\end{equation}
where $\msgf{\mu}{X_{k-1}}$ is the forward sum-product message along the edge $X_{k-1}$
and where $\gamma$ is the missing scale factor in (\ref{eqn:PxCostChain}).

\begin{figure}
\centering
\begin{picture}(75,15)(0,0)
%
\put(0,5){\line(1,0){15}}       \put(7.5,8){\cent{$X_{k-2}$}}
\put(15,2.5){\framebox(5,5){}}  \put(17.5,11){\cent{$g_{k-1}$}}
\put(20,5){\line(1,0){15}}      \put(27.5,8){\cent{$X_{k-1}$}}
\put(35,2.5){\framebox(5,5){}}  \put(37.5,11){\cent{$g_{k}$}}
\put(40,5){\line(1,0){15}}      \put(47.5,8){\cent{$X_{k}$}}
 {\thicklines
  \put(50,3){\vector(-1,0){5.5}} 
 }
\put(55,2.5){\framebox(5,5){}}  \put(57.5,11){\cent{$g_{k+1}$}}
\put(60,5){\line(1,0){15}}      \put(67.5,8){\cent{$X_{k+1}$}}
 {\thicklines
  \put(70,3){\vector(-1,0){5.5}}
 }
\end{picture}
\vspace{-2\unitlength}
\caption{\label{fig:FFGPxCostChainMsgb}%
Forney factor graph of (\ref{eqn:PxCostChain}) 
with messages $\msgb{\mu}{X_k}$~(\ref{eqn:SumProdMsgbXk}).}
\end{figure}

We also note that 
\begin{equation}
\sum_{x_1} \msgb{\mu}{X_1}(x_1) = \sum_{\x} g(\x),
\end{equation}
where $g(\x)$ is defined as the right-hand side of~(\ref{eqn:PxCostChain}). 
In this paper, this fact is used to compute the marginals (\ref{eqn:SampleMarginal})
as a by-product of the sampling.

The generalization of all this to arbitrary factor graphs without cycles 
is straightforward.

\vspace{6ex}


\newcommand{\IT}{IEEE Trans.\ Inf.\ Theory}
\newcommand{\CASI}{IEEE Trans.\ Circuits \& Systems~I}
\newcommand{\COM}{IEEE Trans.\ Comm.}
\newcommand{\COMLet}{IEEE Commun.\ Lett.}
\newcommand{\COMMag}{IEEE Communications Mag.}
\newcommand{\ETT}{Europ.\ Trans.\ Telecomm.}
\newcommand{\SPMag}{IEEE Signal Proc.\ Mag.}
\newcommand{\ProcIEEE}{Proceedings of the IEEE}
\newcommand{\PH}{Phys.\ Rev.\ Lett.}
\newcommand{\Mag}{IEEE Trans.\ Magnetics}


\end{document}